\pdfoutput=1
\documentclass[conference]{IEEEtran}
\ifCLASSINFOpdf
  \usepackage[pdftex]{graphicx}
\else
  \usepackage[dvips]{graphicx}
\fi
\usepackage[cmex10]{amsmath}
\usepackage{array}
\usepackage{url}

\usepackage{cite}
\usepackage{caption}
\usepackage{amsthm}
\usepackage{amsmath}
\usepackage{amsfonts}

\begin{document}

%
\title{Isoform reconstruction using short RNA-Seq reads by maximum likelihood is NP-hard}

\author{
\IEEEauthorblockN{Tianyang Li}
\and
\IEEEauthorblockN{Rui Jiang}
\IEEEauthorblockA{\hspace*{1mm}\\
MOE Key Laboratory of Bioinformatics\\ 
and Bioinformatics Division, \\
TNLIST/Department of Automation, \\
Tsinghua University, \\
Beijing, China, 100084\\
\IEEEauthorrefmark{1}Correspondence: zhangxg@tsinghua.edu.cn}
\and
\IEEEauthorblockN{Xuegong Zhang\IEEEauthorrefmark{1}}
}

\maketitle

\begin{abstract}
Maximum likelihood is a popular technique for isoform reconstruction.  
Here, we show that isoform reconstruction using short RNA-Seq reads by maximum likelihood is NP-hard.
\end{abstract}

\section{Introduction}
Isoform reconstruction is a key step in RNA-Seq analysis. 
Tools such as CEM \cite{Li15112012}, iReckon \cite{Mezlini29112012}, NSMAP \cite{nsmap.21575225}, 
and Montebello \cite{Hiller.Montebello} use maximum likelihood for isoform reconstruction. 
The maximum likelihood approach has been observed to be computationally expensive. 
Here, we show that isoform reconstruction using short RNA-Seq reads by maximum likelihood is NP-hard.

\section{Results}
A Poisson mixture model \cite{Jiang15042009, 2011arXiv1106.3211S, 2011arXiv1104.3889P} is used for isoform reconstruction. 
We represent a gene as a directed acyclic graph $G = (V, E)$ where each vertex in $G$ represents an exon, 
and a path in $G$ represents an isoform of this gene \cite{Heber01072002, scripture.nature}. 
In the model \cite{2011arXiv1104.3889P}, the likelihood of observing $N_s$ at each read 3' end location equivalence class $s$ is 
\begin{equation}
\label{likelihood}
\prod_{s \in S} \frac{e^{-\lambda_s} {\lambda_s}^{N_s}}{N_s !}
\end{equation}
Here $S$ is the set of all read 3' end location equivalence classes, and 
\begin{equation}
\label{loc-expr-level}
\lambda_s = \sum_{i \in I_s} a_{is} \theta_i 
\end{equation}
where $I_s$ is the set of isoforms compatible with $s$, 
$\theta_i$ is isoform $i$'s expression level, 
and $a_{is} > 0$ is the sampling rate for read 3' end location equivalence class $s$ on isoform $i$ \cite{2011arXiv1106.3211S}.

To reconstruct isoforms, 
we seek to maximize the likelihood \eqref{likelihood} with a set $I$ consisting of isoforms, and each isoform's expression level $\theta_i$, $i \in I$. 
In order to explain all observed reads, 
we must be able to align each read to at least one isoform in $I$. 
However, because there are a large number of possible isoforms, 
and it is generally believed that a gene only has a small number of highly expressed isoforms, 
we instead try to find $I$ and $\theta_i$, $i \in I$ that maximize the following penalized likelihood 
\begin{equation}
\label{pen-like}
e^{-K \|I\|_0} \prod_{s \in S} \frac{e^{-\lambda_s} {\lambda_s}^{N_s}}{N_s !} 
\end{equation}
where $\|I\|_0$ is the number of $\theta_i > 0$, $i\in I$, and  $K > 0$ is a real constant. 
Note that setting $k = \frac{1}{2} \log(\sum_{s \in S} N_s)$ is equivalent to using the Bayesian information criterion \cite{BIC.Schwarz_1978} for an equivalent multinomial model \cite{2011arXiv1104.3889P, cufflinks.2010}. 

\newtheorem*{misof}{M-ISOFORM}

To show the hardness of isoform reconstruction by maximizing the penalized likelihood \eqref{pen-like}, 
we consider the following decision problem
\begin{misof}
\hspace*{1mm}

\textnormal{INSTANCE:} 
A set of reads aligned to a gene where the read count at a read 3' end location equivalence class $s$ the read count is $N_s$ .

\textnormal{QUESTION:} 
Does there exist an isoform set $I$ with at most $m$ isoforms such that $\prod_{s \in S} \frac{e^{-\lambda_s} {\lambda_s}^{N_s}}{N_s !} \geq p$?
\end{misof}

\newtheorem{theorem}{Theorem}

\begin{theorem}
\label{misof-npc}
\textnormal{M-ISOFORM} is NP-complete.
\end{theorem}

\section{Discussion}
We can avoid computationally determining a gene's isoform set if laboratory protocols can be used to find the gene's existing isoforms. 
It is possible to use methods such as paired-end tag sequencing \cite{Fullwood01042009}, 
and single-molecule sequencing \cite{hybrid.rna.seq.2012} to determine a gene's isoforms. 
We expect these and related technologies to mature in the foreseeable future, 
reducing the demand for computational resources. 

Haplotype inference \cite{Li_Kim_Waterman_2004, Xing:2004:BHI:1015330.1015423}, a similar problem , is also known to be NP-hard \cite{1668028}. We believe that haplotype inference will also benefit from technologies offering longer sequencing reads.

\section{Proof}
The proof borrows ideas from \cite{Tomescu.recomb.seq.2013}, 
where a network flow approach is used for isoform reconstruction. 
To show that M-ISOFORM is is NP-complete,  
we reduce 3-PARTITION \cite{Garey:1990:CIG:574848, doi:10.1137/0204035}, 
a strongly NP-complete problem, to M-ISOFORM. 
We use an approach similar to the one used in \cite{Vatinlen20081390}, 
where flows are split into paths. 

\begin{IEEEproof}[Proof of Theorem \ref{misof-npc}]
\newtheorem*{part3}{3-PARTITION}
3-PARTITION is stated in \cite{Garey:1990:CIG:574848} as follows
\begin{part3}
\hspace*{1mm}

\textnormal{INSTANCE:} 
Set $X$ of $3 w$ elements, a bound $Y \in \mathbb{Z}^+$, 
and a size $u(x) \in \mathbb{Z}^+$ for each $x \in X$ such that $\frac{Y}{4} < u(x) < \frac{Y}{2}$ and such that $\sum_{x \in X} u(x) = w Y$.

\textnormal{QUESTION:} 
Can $X$ be partitioned into $w$ disjoint sets $X_1$, $X_2$, ..., $X_w$ such that, 
for $1 \leq i \leq w$, $\sum_{x \in X_i} u(x) = Y$ 
(note that each $X_i$ must therefore contain exactly three elements from $X$)?
\end{part3}

\begin{figure}[!t]
\centering
\includegraphics[width=0.5\textwidth]{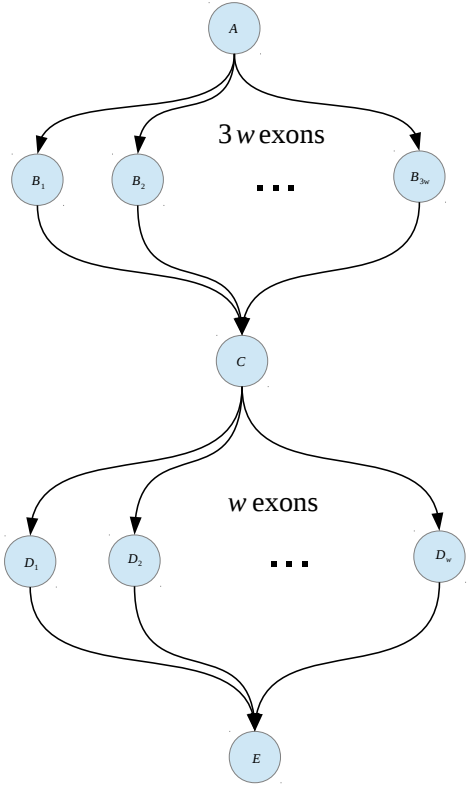}
\caption{A gene structure for an instance of 3-PARTITION}
\label{part3-gene}
\end{figure}

For an instance of 3-PARTITION, we create a gene with structure as shown in Figure \ref{part3-gene}. 
Let $E \in \mathbb{Z}^+$ be a fixed constant, we make each exon $R$ bp long, 
and each read $R + 1$ bp long. 
At each bp in exon $A$, 
for $1 \leq i \leq 3 w$ we have $u(x_i)$ reads starting at this location going to exon $B_i$, 
thus there are $\sum_{1 \leq i \leq 3w} u(x_i)$ reads starting at this location. 
For $1 \leq i \leq 3 w$, at each bp in exon $B_i$, we have $u(x_i)$ reads starting at this location going to exon $C$. 
At each bp in exon $C$, 
for $1 \leq i \leq w$ we have $Y$ reads starting at this location going to exon $D_i$, 
thus there are $z Y$ reads starting at this location. 
For $1 \leq i \leq w$, at each bp in exon $D_i$, we have $Y$ reads starting at this location going to exon $E$. 
We will show that this instance of 3-PARTITION has a solution if and only if there exists an isoform set $I$ with at most $3 z$ isoforms such that $\prod_{s \in S} \frac{e^{-\lambda_s} {\lambda_s}^{N_s}}{N_s !} \geq \prod_{s \in S} \frac{e^{-N_s} {N_s}^{N_s}}{N_s !}$.

It is easy to see that $\prod_{s \in S} \frac{e^{-\lambda_s} {\lambda_s}^{N_s}}{N_s !} \geq \prod_{s \in S} \frac{e^{-N_s} {N_s}^{N_s}}{N_s !}$ if and only if $\forall s \in S$, $\lambda_s = N_s$. 

If we have a solution to this instance of 3-PARTITION, 
it is easy to verify that an isoform set with $3 z$ isoforms where $\theta_i = u(x_i)$, 
$1 \leq i \leq 3 z$, and isoform $i$ consists of exons $A$, 
$B_i$, $C$, $D_j$ ($j$ satisfies $x_i \in A_j$), 
and $E$ is a solution to this particular instance of M-ISOFORM.

If we have a solution to this particular instance of M-ISOFORM, 
we show that there also exists a solution to the instance of 3-PARTITION. 
In this case, the isoform set must have exactly $3 z$ isoforms, 
because at least $3 z$ isoforms are required to explain all the reads. 
Thus, for $1 \leq i \leq 3 z$ we have $\theta_i = u(x_i)$. 
Because we also have $\forall s \in S$, $\lambda_s = N_s$, 
and $\forall x \in X$, $\frac{Y}{4} < u(x) < \frac{Y}{2}$
we can see that for $1 \leq j \leq w$ exon $D_j$ has exactly three isoforms passing through it, 
and $\sum_{\text{$i$ passes through $D_j$}} \theta_i = Y$. 
Therefore, we have a solution to the instance of 3-PARTITION. 

It is easy to see that M-ISOFORM is in NP if we use the real RAM model \cite{Preparata:1985:CGI:4333}. 
Because 3-PARTITION is strongly NP-complete \cite{Garey:1990:CIG:574848, doi:10.1137/0204035}, 
we conclude that M-ISOFORM is NP-complete. 

\end{IEEEproof}

\section*{Acknowledgments}
We thank Feng Zeng for insightful discussions. 

\bibliographystyle{IEEEtran}
\bibliography{IEEEabrv,mybib}

\end{document}